\documentstyle[preprint,aps]{revtex}
\input{psfig}

\begin{document}

\title{Unconditional Continuous Variable Dense Coding.}
\author{T.C.Ralph}
\address{Centre for Quantum Computer Technology,
\\ Department of Physics, The University of Queensland, 
\\ St Lucia 4072 Australia \\
 E-mail: ralph@physics.uq.edu.au}
 \author{E.H.Huntington}
\address{School of Electrical Engineering, \\
University College, \\
University of New South Wales, \\
Australian Defence Force Academy \\
Canberra ACT 2600 Australia}
\maketitle

\begin{center}
\scriptsize (February 2002)
\end{center}

\begin{abstract}

We investigate the conditions under which unconditional dense coding
can be achieved using continuous variable entanglement. We consider
the effect of entanglement impurity and detector efficiency and
discuss experimental verification. We conclude that the requirements
for a strong demonstration are not as stringent as previously thought
and are within the reach of present technology.

\end{abstract}

\vspace{10 mm}

\section{Introduction}

The classical channel capacity of a quantum channel can be enhanced
if the sender and recipient of the information, Alice and Bob
repectively, share an entangled state. This effect is known as quantum
dense coding \cite{ben92} and can be thought of as the converse
problem to quantum teleportation \cite{ben93} where, effectively, the
quantum capacity of a classical channel is enhanced by the use of
entanglement.

Dense coding was originally introduced for discrete variables and an
experimental demonstration of the effect has been made using
photonic polarization entanglement \cite{mat96}. One drawback of this
demonstration was, due to the low efficiency of entanglement
production and detection, the demonstration was conditional on Bob
detecting a pair of photons, a rare event. In contrast a dense coding
scheme based on continuous variables, such as the quadrature
amplitudes of a light field, which has recently been proposed, would
in principle demonstrate an unconditional improvement in classical
channel capacity \cite{bra99}. Ultimately this scheme can beat, under certain
conditions, the maximum channel capacity given by Fock state encoding.
However the conditions for this strong violation found in \cite{bra99}
required
unrealistic levels of squeezing.

A number of groups have taken steps
towards the experimental implementation of this scheme
\cite{kim99,peng00}. In these experiments increased signal to noise
was demonstrated with the addition of entanglement and conclusions
were drawn about the violation of coherent state classical capacity,
based on the results of Ref \cite{bra99}. However unit
entanglement state purity and
detection efficiency were assumed in \cite{bra99}, which is unlikely
to have been the case experimentally. Also no attempt was made to
experimentally quantify the number of quanta used in the communication
channel. There is thus a need for a more
detailed analysis.

In this paper we make such an investigation. We come to the rather
surprizing conclusion that in fact the conditions required for a
strong demonstration of the effect, i.e. beating the
ultimate channel capacity given by Fock state encoding, are not as
stringent as previously thought, even taking into account lack of
state purity and non-unit detection efficiency.

\section{Ideal Channel Capacities}

\begin{figure}
   \begin{center}
   \begin{tabular}{c}
   \psfig{figure=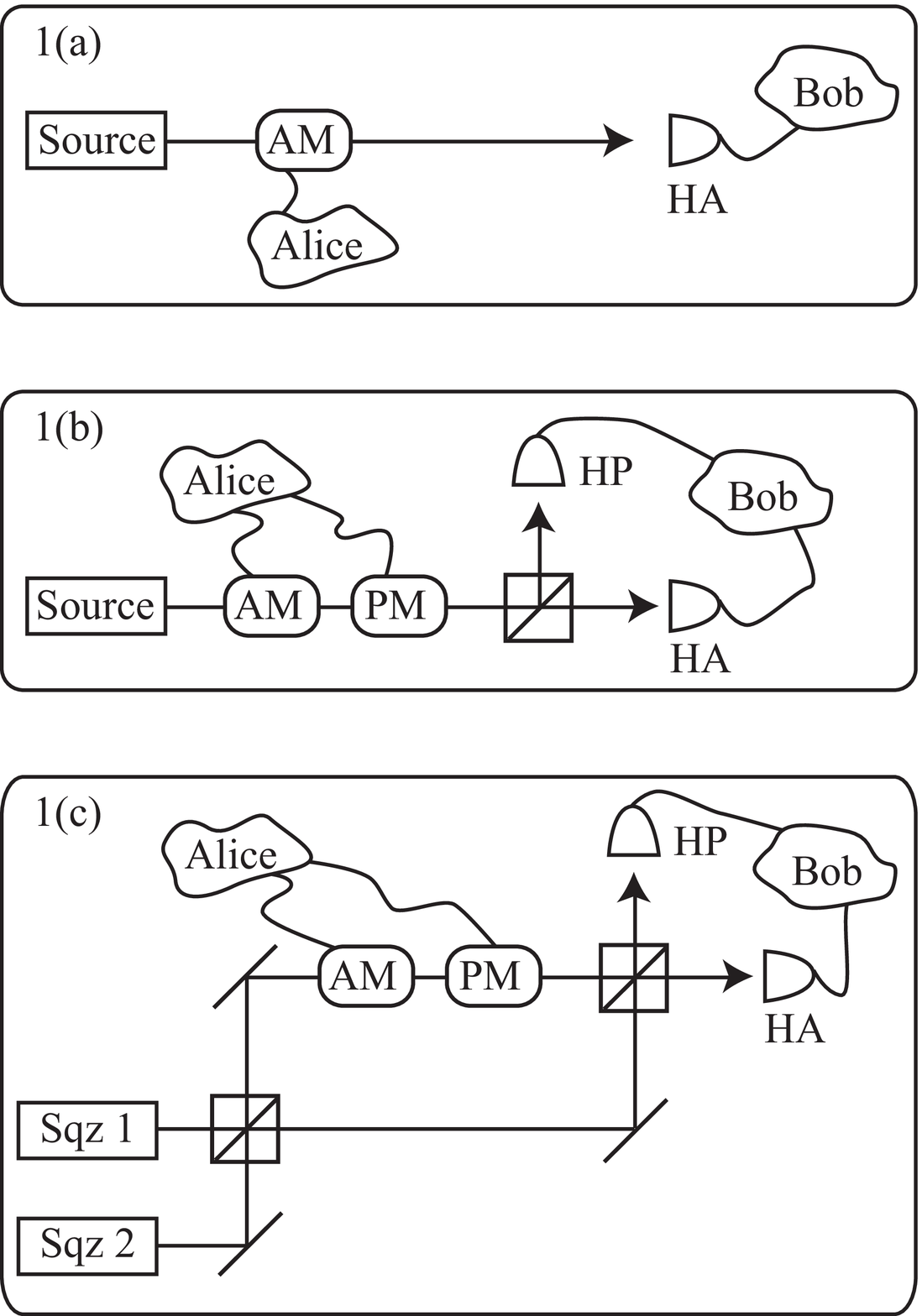,width=10cm}
   \end{tabular}
   \end{center}
 \caption{\label{fig1}  Schematic diagrams of (a) coherent homodyne (b)
 coherent heterodyne (c) dense coding schemes for a communication
 channel.  The abbreviations are: AM = amplitude modulation; PM =
 phase modulation; HA = coherent homodyne detection of the amplitude
 quadrature; HP = coherent homodyne detection of the phase
 quadrature.  The beamsplitters are taken to be 50 \%
 transmitting and the two squeezed sources are squeezed in orthogonal
 quadratures.}
\end{figure}

We begin by rederiving the channel capacities of Gaussian quantum
channels and continuous variable dense coding using quadrature
spectral variances. Such variances are directly measureable in an
experiment.
The Shannon capacity \cite{sha48} of a communication channel
with Gaussian noise of power (variance) $N$ and Gaussian distributed
signal power $S$ operating at the bandwidth limit is
\begin{equation}
C={{1}\over{2}} \log_{2}[1+{{S}\over{N}}]
\label{cc}
\end{equation}
Eq.\ref{cc} can be used to calculate the channel capacities of quantum
states with Gaussian probability distributions such as coherent states
and squeezed states \cite{yam86,cav94}. Consider first a signal composed of a
Gaussian distribution of coherent state amplitudes all with the
same quadrature angle (see Fig.\ref{fig1} (a)).
The signal power $V_{s}$ is given by the variance
of the distribution. The noise is given by the intrinsic quantum
noise of the coherent states and is defined to be $V_{n}=1$.
Because the quadrature angle of the signal is known, homodyne
detection can in principle detect the the signal without further
penalty thus the measured signal to noise is $S/N=V_{s}/V_{n}=V_{s}$.

In general the average photon number per bandwidth per second
of a light beam is given
by
\begin{equation}
\bar n = {{1}\over{4}}(V^{+}+V^{-})-{{1}\over{2}}
\label{n}
\end{equation}
where $V^{+}$ ($V^{-}$) are the variances of the maximum (minimum)
quadrature projections of the noise ellipse of the state. These
projections are orthogonal quadratures, such as amplitude and phase,
and obey the uncertainty principle $V^{+}V^{-} \ge 1$. In the above
example one quadrature is made up of signal plus quantum
noise such that $V^{+}=V_{s}+1$ whilst the orthogonal quadrature is just
quantum noise so $V^{-}=1$. Hence $\bar n= 1/4 V_{s}$ and so the
channel capacity of a coherent state with single quadrature encoding
and homodyne detection is
\begin{equation}
C_{c}=\log_{2}[\sqrt{1+4 \bar n}]
\label{cccho}
\end{equation}
Establishing in an experiment that a particular optical mode has this
capacity would involve: (i) measuring the quadrature
amplitude variances of the beam, $V^{+}$ and $V^{-}$,
(ii) calibrating Alice's signal
variance and (iii) measuring Bob's signal to noise. If these
measurments agreed with the theoretical conditions above then
Shannons theorem tells us that an encoding scheme exists which could
realize the channel capacity of Eq.\ref{cccho}. An example of such an
encoding is given in \cite{enc}.

For photon numbers $\bar n > 2$ improved channel capacity can be
obtained by encoding symmetrically on both quadratures and detecting both
quadratures simultaneously using heterodyne detection or dual homodyne
detection (see Fig.\ref{fig1}(b)). Because of
the non-commutation of orthogonal quaratures there is a penalty for
their simultaneous detection which reduces the signal to noise of each
quadrature to $S/N=1/2 V_{s}$. Also because there is signal on
both quadratures the average photon number of the beam is
now $\bar n= 1/2 V_{s}$. On the other hand the total
channel capacity will now be the sum of the two independent channels
carried
by the two quadratures. Thus the channel capacity for a coherent
state with dual quadrature encoding and heterodyne detection is
\begin{eqnarray}
 C_{ch} & = & {{1}\over{2}}
\log_{2}[1+{{S}\over{N}}^{+}]+{{1}\over{2}}
\log_{2}[1+{{S}\over{N}}^{-}] \nonumber\\
& = & \log_{2}[1+\bar n]
\label{ccchet}
\end{eqnarray}
which exceeds that of the homodyne technique (Eq.\ref{cccho}) for $\bar n>2$.

The above channel capacities are the best achievable if we restrict
ourselves to a semi-classical treatment of light. However the  channel
capacity
of the homodyne technique (Fig.\ref{fig1}(a)) can be improved by the use of
non-classical,
squeezed light. With squeezed light the noise variance of the encoded
quadrature
can be reduced such that $V_{ne}<1$, whilst the noise of the unencoded
quadrature is increased such that $V_{nu} \ge 1/V_{ne}$. As a result
the signal to noise is improved to $S/N=V_{s}/V_{ne}$ whilst the
photon number is now given by Eq.\ref{n} but with
$V^{+}=V_{s}+V_{ne}$ and $V^{-}=1/V_{ne}$ where a pure (i.e.
minimum uncertainty) squeezed state has been assumed. Maximizing the signal
to noise for fixed $\bar n$ leads to $S/N=4(\bar n+\bar n^{2})$ for a
squeezed quadrature variance of $V_{ne,opt}=1/(1+2 \bar n)$.
Hence the channel capacity for a squeezed beam with homodyne
detection is
\begin{equation}
C_{sh}=\log_{2}[1+2 \bar n]
\label{ccsho}
\end{equation}
which exceeds both coherent homodyne and heterodyne for all values
of $\bar n$.

A final improvement in channel capacity can be obtained by allowing
non-Gaussian states. The absolute maximum channel capacity for a
single mode is given by the Holevo bound and can be realized by
encoding in a maximum entropy ensemble of Fock states and
using photon number detection \cite{yam86,cav94,yuen93}.
This ultimate channel capacity is
\begin{equation}
C_{Fock}=(1+ \bar n) \log_{2} [(1+\bar n)]-\bar n \log_{2} [\bar n]
\label{ccf}
\end{equation}
which is the maximal channel capacity at all values of $\bar n$.

We now turn to dense coding. At low average photon numbers the single 
channel capacities are always best. However, we will find that for 
sufficiently high average photon numbers dense coding can give 
superior capacities. The set-up is depicted in Fig.\ref{fig1}(c).
Entanglement is generated in the standard way by mixing two squeezed
states,
with their squeezing ellipses orthogonal,on a 50:50
beamsplitter \cite{yeo93}. One half of the entangled pair is sent to
Alice who encodes on both quadratures in the manner of coherent
heterodyne. She sends the beam on to Bob who has also received the
other half of the entangled pair. He uses a dual homodyne technique to
measure both quadratures of the beam from Alice but injects his
entangled beam into the empty port of the dual homodyne beamsplitter.
The resulting signal to noise for the two quadrature channels is
$S/N=1/2 (V_{s}/V_{ne})$, where now $V_{ne}$ is the variance of the
squeezed quadrature of the beams used to create the entanglement. The
photon number is just that of the beam carrying the signal (the
cost of distributing the entanglement is not taken into account) and
so is given by Eq.\ref{n} with $V^{+}=1/2 V_{s}+1/4 V_{ne}$ and
$V^{-}=1/V_{ne}$. Once again pure squeezed states are assumed.
Maximizing the signal to noise for fixed $\bar n$ gives $S/N=\bar
n+\bar n^{2}$ for a squeezed quadrature variance of $V_{ne,opt}=1/(1+2 \bar
n)$.
So the optimum channel capacity for dense coding is
\begin{equation}
C_{dc}^{opt}=\log_{2}[1+\bar n+\bar n^{2}]
\label{ccdc}
\end{equation}
which exceeds the coherent state homodyne for $\bar n > 0.478$, which
can be achieved with $V_{ne} \approx 0.5$ (about 50\% squeezing), and
always exceeds the coherent state heterodyne channel capacity.  Dense
coding beats the squeezed state channel capacity with $\bar n > 1$
(achieved with $V_{ne} \approx 0.33 $ or about 67\% squeezing) and
beats Fock state encoding when $\bar n> 1.88$ (achieved with
$V_{ne}\approx 0.2$ or squeezing of about 80\%).

Some comments are in order concerning the analysis to date.  Firstly
notice the boundaries of the previous analysis were for pure squeezed
states which saturate the uncertainty inequality.  In contrast the
states produced in experiments are rarely pure, sometimes because of
technical noise \cite{sol}, sometimes due to the type of squeezing
mechanism \cite{dio}, and sometimes simply due to loss in the
non-linear crystal \cite{buc00}. Loss in the optical elements 
used to produce the entanglement from the squeezing will also reduce 
the purity (as well as the effective entanglement). 
Therefore, in an experiment we will
have that $V_{nu}=1/V_{ne}+b$, where $b$ represents excess noise.
This means that a particular level of entanglement is accompanied by
more photons than in the pure case.  Hence channel capacities will be
lowered\footnote{For example, the optimum signal to noise ratio of the
dense coding scheme when considering entanglement impurity is
$S/N=\bar n+\bar n^{2}-b(0.25+0.5 \bar n)+0.0625 b^{2}$.  At the
optimum squeezed quadrature variance of $V_{ne,opt}=2/(4 \bar n+2-b)$
(with $b< 4 \bar n+2$) the new, more general expression for the
optimum dense coding capacity is $C_{dci}^{opt}=\log_{2}[1+\bar n+\bar
n^{2}-b(0.25+0.5 \bar n)+0.0625 b^{2}]$ which will be less
than $C_{dc}^{opt}$ for any amount of excess noise.}.  Further, unit
detection efficiency was assumed.  Again, this is unlikely in an
experiment.  As a result Bob's detected variances will be given by
$V_{det}=\eta V + 1-\eta$, where $\eta$ is the detection efficiency.
Non-unit detection efficiency will lower signal to noise and once
again decrease the effective channel capacity. Propagation loss 
(assumed equal in the two channels) has 
the same effect as detection efficiency and so can be rolled into the 
value of $\eta$.

To achieve {\it unconditional dense coding} we require that even in
the presence of these kinds of imperfections, the dense coding channel
capacity exceeds that of the ideal single channel capacities.  The
levels of squeezing apparently required in the ideal case are already
at the boundary of what is currently achievable experimentally -
experiments regularly achieve squeezing greater than 3dB (50\%)
\cite{sqvar} but stable measured squeezing of approximately 5dB
(~68\%) has only been reported recently \cite{buc01}.  Imperfections
appear to only further increase the stringent experimental
requirements and it would seem that an experimental
demonstration of unconditional dense coding is beyond current
technology.  However, in the following section we will show that this
is not the case and that a demonstration is within the reach of
current continuous variable technology.

 \section{Demonstrating Unconditional Dense Coding}

 Notice that the preceeding analysis and that of Ref \cite{bra99}
 asked the question: ``what is the minimum {\it photon number} for which we
 can demonstrate dense coding''. We will now show that a different
 answer is obtained if we ask the question: ``what is the minimum {\it
 squeezing} required to demonstrate dense coding''. Rather than 
 maximizing the signal to noise ratio for a fixed $\bar{n}$ we now 
 allow an arbitrary relationship between the squeezed quadrature 
 variance, average photon number and excess noise. The detected dense 
 coding signal
to noise
 ratio may then be explicitly written as 
 $(\eta(4 \bar n-V_{ne}-1/V_{ne}-b+2))/(4 \eta
 V_{ne}+4-4\eta)$.  Hence a more general expression for the dense coding
capacity is

\begin{equation}
 C_{dc}= \log_{2}\left[1+\frac{\eta(4 \bar
 n-V_{ne}-1/V_{ne}-b+2)}{4 (\eta V_{ne}+1-\eta)} \right]
    \label{ccdciarb}
\end{equation}

\begin{figure}
   \begin{center}
   \begin{tabular}{c}
   \psfig{figure=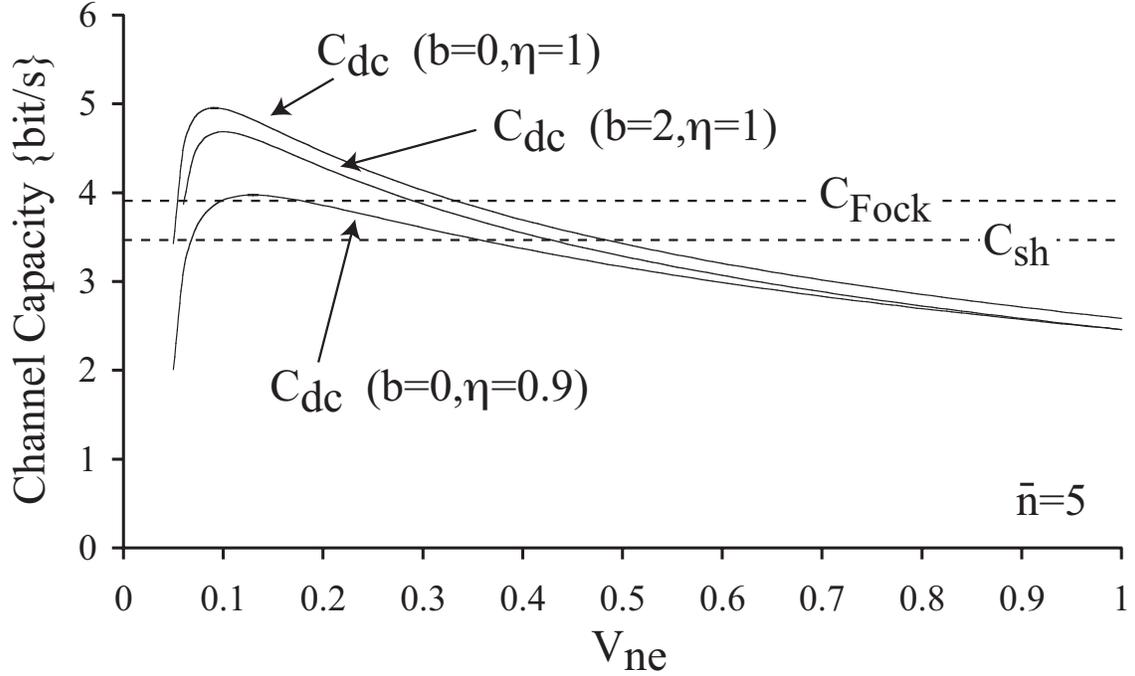,width=15cm}
   \end{tabular}
   \end{center}
 \caption{\label{fig2} Plots of the dense coding channel capacity,
 $C_{dc}$, as a function of input squeezing, $V_{ne}$, for an average
 photon number of $\bar n=5$.  Regions in which the dense coding
 channel capacity exceed the Fock state channel capacity display
 unconditional dense coding with respect to the Fock state scheme.
 Regions in which $C_{dc}>C_{sh}$ display unconditional dense coding
 with respect to the squeezed state scheme.}
\end{figure}

Fig.\ref{fig2} shows the channel capacity of the dense coding scheme,
$C_{dc}$ as a function of the squeezed quadrature variance at an
average photon number of $\bar n = 5$.  For the moment, focus on the
topmost curve (labelled ``$C_{dc}$ $(b=0,\eta=1)$'') which
represents the channel capacity in the absence of excess noise and for
perfect detection efficiency.  This curve represents the best possible
scenario for that photon number.  Also shown in this Figure are the
ideal maximum channel capacities for the squeezed state scheme,
$C_{sh}$ and
the Fock state scheme $C_{Fock}$ for this photon number.  As
illustrated in Fig.\ref{fig2} and indicated by Equation \ref{ccdc}, at
$\bar n = 5$ the optimum dense coding capacity exceeds the capacity of
the squeezed or Fock state schemes.  However, Fig.\ref{fig2}
illustrates a point that is not clear from Equation \ref{ccdc} - the
dense coding channel capacity exceeds that of $C_{sh}$ or $C_{Fock}$
for a number of values of $V_{ne}$, not just the optimum.

Most
significantly, Fig.\ref{fig2} shows that it is possible to demonstrate
unconditional dense coding with a relatively modest amount of
squeezing.  For example, it is seen from Fig.\ref{fig2} that
unconditional
dense coding may be demonstrated with respect to the squeezed state
system for $V_{ne} \approx 0.48$ and it may be demonstrated with
respect to the Fock state scheme for $V_{ne} \approx 0.33$.  These
levels of squeezing are far more experimentally feasible than those
found by simply considering the optimum.  Even more heartening from an
experimental perspective is that the levels of squeezing required may
be reduced by increasing $\bar n$.  Experimentally, given a minimum
amount of squeezing, this amounts to simply increasing the signal
strength.

\begin{figure}
   \begin{center}
   \begin{tabular}{c}
   \psfig{figure=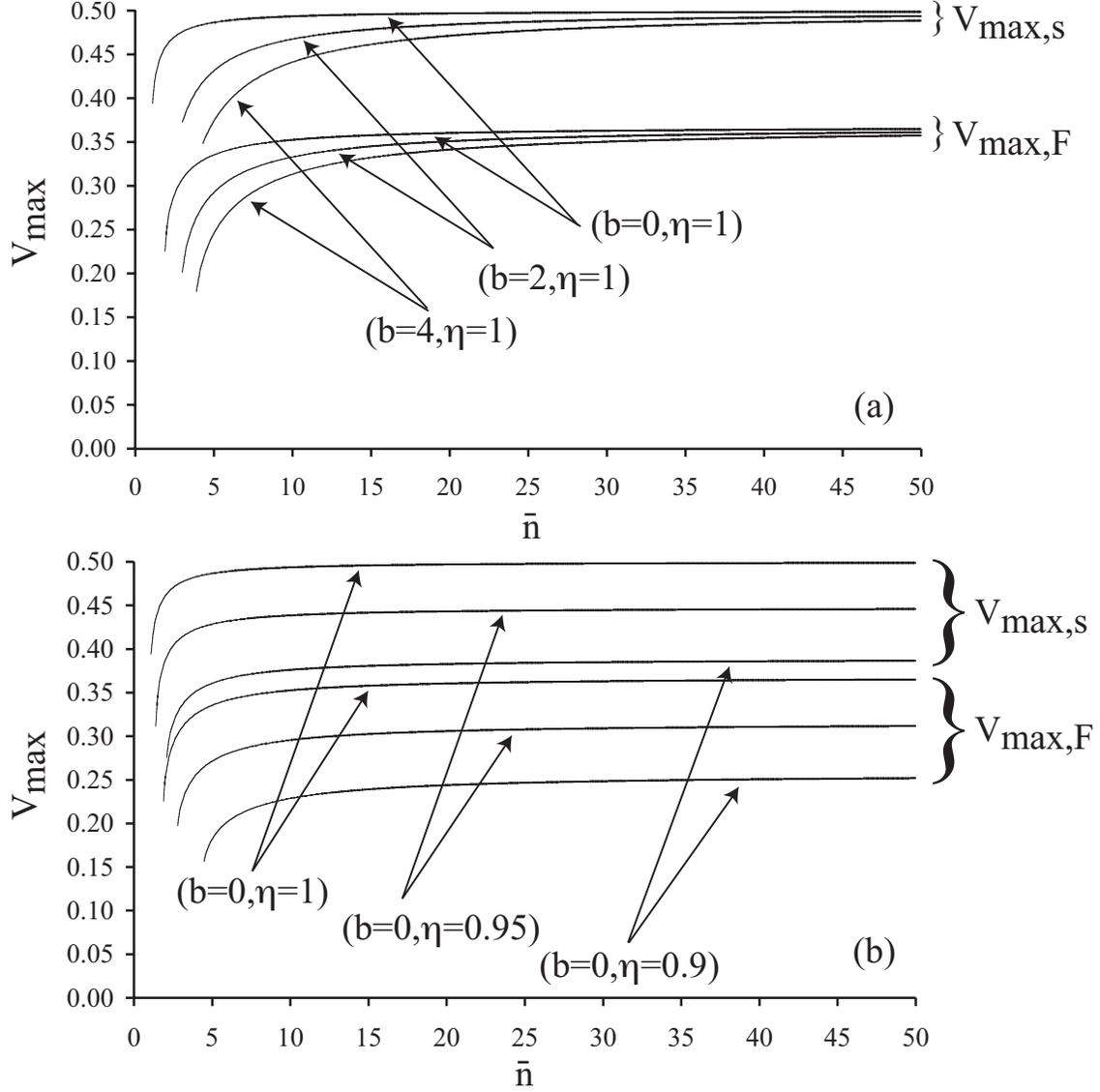,width=15cm}
   \end{tabular}
   \end{center}
\caption{\label{fig3}  Plots of the maximum squeezed quadrature
variances at which unconditional dense coding may be demonstrated with
respect to $C_{sh}$ and $C_{Fock}$.  These are labelled $V_{max,s}$
and $V_{max,F}$ respectively.}
\end{figure}

Denoting the maximum squeezed quadrature
variances at which unconditional dense coding may be demonstrated with
respect to $C_{sh}$ and $C_{Fock}$ as $V_{max,s}$ and $V_{max,F}$
respectively, Fig.\ref{fig3} shows $V_{max,s}$ and $V_{max,F}$ as a
function of the photon number.  Again, focussing for the moment on the
curves for pure entanglement and perfect detection efficiency
(labelled ``($b=0,\eta=1$)'', it is seen that $V_{max,s}$ and
$V_{max,F}$ asymptote to values of $1/2$ and $1/e$ respectively.  This
is quite a surprizing result.  This Figure shows that, in a perfect
experiment, it is possible to demonstrate unconditional dense coding
with 50 \% squeezing with respect to the the squeezed state
channel capacity and 63 \% squeezing with respect to the Fock state
scheme.

Turning now to experimental issues such as excess noise or imperfect
detection efficiency, Fig.\ref{fig2} also shows curves representing
the dense coding channel capacity, $C_{dc}$ when excess noise,
labelled ``($b=2,\eta=1$)'' and imperfect detection efficiency, labelled
``($b=0,\eta=0.9$)'', are considered.  First note that Fig.\ref{fig2}
shows that both excess noise and imperfect detection efficiency
decrease the effective channel capacity of the dense coding scheme.
Indeed, sufficient amounts of either of these imperfections will
render a demonstration of unconditional dense coding impossible for
low photon numbers.  However, given a minimum level of squeezing, this
may be solved by increasing the photon number, i.e.
by increasing the signal strength.  The
minimum level of squeezing required for each photon number depends
quite strongly on the entanglement impurity and detection efficiency.
This effect is shown in Fig.\ref{fig3} where $V_{max,F}$ and 
$V_{max,s}$ are plotted
as a function of the photon number for a number of values of 
excess noise in part (a) and a number of values of detection 
efficiency in part (b).

Fig.\ref{fig3} shows that, for a given photon number, either of these
imperfections will mean that more squeezing is required than for the
best possible scenario. Alternatively, for a given level of squeezing,
entanglement impurity or imperfect detection efficiency will require
that more photons must be used to demonstrate unconditional dense
coding.

Perhaps more importantly, Fig.\ref{fig3} shows that the experimental
issue of greatest concern is that of imperfect detection efficiency.
Excess noise will certainly have an effect on the performance of the
dense coding scheme for small photon numbers.  However the asymptotes
of $V_{max,F}$ and $V_{max,s}$ do not depend on the entanglement
purity.  By contrast, the asymptotes of $V_{max,F}$ and $V_{max,s}$
depend very strongly on the detection efficiency.  In practical terms,
this means that much greater levels of squeezing will be required to
demonstrate unconditional dense coding when the detection efficiency
is poor. Indeed the amount of squeezing required increases exponentially 
around a
characteristic value of detection efficiency.  This suggests that practical
systems must exceed a minimum detection efficiency in order to demonstrate
unconditional dense coding.   When comparing the dense coding channel
capacity to the squeezed state system, the minimum detection efficiency
required is $\eta_{min,s}=2/3$.  The minimum detection efficiency increases
to $\eta_{min,F}= e/(1+e)$ in order to demonstrate unconditional dense
coding with respect to the Fock state system. 

Taking these effects into account, it appears that an experimental
demonstration of unconditional dense coding with respect to the
squeezed state system is currently feasible.  For example, with
experimentally realistic detection efficiencies of 85-95 \% the
maximum squeezing required would be approximately 68-55 \%
respectively.  These levels of squeezing are now quite commonly
achieved \cite{sqvar,buc01}.  On the other hand, an experimental
demonstration of unconditional dense coding with respect to the Fock
scheme is a rather more ambitious, but not unattainable, goal.  With
experimentally realistic detection efficiencies of 85-95 \% the
maximum squeezing required would be approximately 81-68 \%
respectively.  Ref.  \cite{buc01} reported measured squeezing of 5dB
with a detection efficiency of approximately 87 \%. Assuming 
a dense coding scheme with
approximately the same detection efficiency as quoted in reference
\cite{buc01}, we conclude that this would have been just sufficient to
demonstrate unconditional dense coding with respect to the Fock state
scheme.  Thus the levels of squeezing required for a strong
demonstration of unconditional dense coding are within the boundaries
of current technology.

\section{Conclusion}

We have shown that by working in the large signal regime
a demonstration of unconditional dense coding appears possible with
present technology. We believe that such a demonstration would
represent a bench-mark experiment in continuous variable quantum
information technology. It is interesting to note relationships
between the entanglement requirements of dense coding and
teleportation. Beating the coherent state channel capacity with dense
coding can be achieved with any level of squeezing in the entanglement.
Similarly an improvement over the classical fidelity limit for
teleportation of coherent states is achieved with any finite level of
squeezing. However the preservation of non-classical properties of
the state like squeezing requires greater than 50\% squeezing in
teleportation, corresponding to the requirement of 50\% squeezing to
beat the squeezed state channel capacity in dense coding. It is
interesting to muse as to whether
the $1/e$ entanglement requirement for
unconditional dense coding corresponds to the passing of some other
tangible limit in teleportation.

\acknowledgments
We thank C.Savage and W.Bowen for useful comments. 
This work was supported by the Australian Research Council.

\newpage

\newpage

\newpage

\end{document}